\documentclass[12pt]{spieman}  
\usepackage{amsmath,amsfonts,amssymb}
\usepackage{graphicx}
\usepackage{setspace}
\usepackage{tocloft}

\title{Limits on Wave Optics Simulations of Plane Wave Propagation in Non-Kolmogorov Turbulence.}

\author[a]{Jeremy P. Bos}
\author[a]{Stephen Grulke}
\author[a]{Jeff Beck}
\affil[a]{Michigan Technological University, Department of Electrical and Computer Engineering, 1400 Townsend Dr., Houghton, MI USA}

\cftpagenumbersoff{figure}
\cftpagenumbersoff{table} 
\begin{document} 
\maketitle

\begin{abstract}
We derive limits for wave optics simulations of plane wave propagation in non-Kolmogorov turbulence using the split-step method and thin phase screens. These limits are used to inform two simulation campaigns where the relationship between volume turbulence strength and normalized intensity variance for various non-Kolmogorov power-laws. We find that simulations of smaller power laws are limited turbulence strengths with Rytov numbers of 7 when the simulation side-length sampling rate is 8192. Under these same conditions it is possible to simulate volume turbulence strength out to Rytov numbers of 12 for Kolmogorov power-laws and higher. We show that the peak scintillation and Rytov number where peak scintillation occur increases monotonically with power-law. Also, that if turbulence strength is fixed, the relationship between scintillation index and power-law depends on the operating regime. In weak turbulence the relationship is negative, and it is positive in stronger turbulence. This work also emphasizes the importance of properly scaling turbulence strength with comparing results with different power-laws and the influence and importance of defining inner and outer scales in these simulations. 
\end{abstract}


{\noindent \footnotesize\textbf{*}contact author,  \linkable{jpbos@mtu.edu} }

\begin{spacing}{2}   
\section{Introduction}
The propagation of optical waves in random media, including atmospheric turbulence, is governed by the stochastic Hemholtz equation. Applying weak perturbation theory, it is possible to achieve closed form expressions for most field statistics as outlined by Tatarskii\cite{tatarski1967wave}. Similarly, asymptotic theory\cite{hill1981theory},\cite{hill1982theory}, provides an excellent match to experiment when fluctuations are very large. In between these two regions Extended Rytov Theory\cite{andrews1999theory} (ERT) provides a heuristic analytical model that has found a good match to some experimental data. Attempts at providing solutions in this region that do not rely on approximations involve complex multi-dimensional integrals and are difficult to generalize\cite{Barakat:99},\cite{flatte1987path},\cite{dashen1979path},\cite{dashen1984distribution}. 

While not analytical, Wave Optics Simulations (WOS) are known to be an accurate approximation to the stochastic Hemholtz equation\cite{flatte2000irradiance}. This approximation is achieved by discretizing the propagation volume and collapsing the media in each segment into a thin phase screen. The optical field is then propagated between adjacent segments via a Fresnel propagation operator. The relative fluctuations at each screen are weak such that the approximation is equivalent to the small perturbation approximation used by Tatarskii. The main strength of WOS is that they can provide insight into wave propagation statistics and the performance of systems that attempt to compensate for the effects of the media and are relatively straightforward to execute.

Though the technique was previously described by others, relative to optical wave propagation Martin and Flatt\'e \cite{martin1988intensity} were the first to explore the use WOS for optical waves. Their studies \cite{martin1988intensity} of plane and spherical waves\cite{martin1990simulation} in random media described by a power-law examines the intensity scintillation as a function of fluctuation strength in the media. Since that time WOS have become the default tool used in modeling the performance of adaptive optics, beam projection, and Free Space Optical (FSO) communication systems \cite{schmidt2010numerical},\cite{voelz2011computational},\cite{belmonte2000feasibility},\cite{johansson1994simulation},\cite{frehlich2000simulation}. Simulations involving phase screens are also commonly used to create synthetic image data of scenes blurred by turbulence \cite{Hardie2017},\cite{bos2012technique},\cite{carrano2003anisoplanatic}. Until relatively recently it has been common to model atmospheric turbulence in WOS as purely Kolmogorov with an infinite outer scale and zero inner scale. 

In 1994 Dauldier \cite{dalaudier1994direct} published evidence of turbulence in the upper atmosphere from balloon-borne measurements described as non-Kolmogorov in nature. Soon after Beland \cite{beland1995some} examined its implications to existing turbulence theory. Strilbing\cite{stribling1995optical} then extended those results to weak fluctuation theory. In this context, non-Kolmogorov turbulence refers to deviation of the power-law away from the 2/3 power-law structure function description of turbulence predicted by Kolmogorov in the inertial subrange between the energy input region (outer-scale) and dissipation region (inner scale). The equivalent two-dimensional energy spectral density of turbulent index of refraction fluctuations shows an -11/3 slope as a function of increasing spatial frequency. This latter quantity is more often found in modern analytical models that begin with the refractive index fluctuation energy spectrum starting with Toselli\cite{toselli2007scintillation},\cite{Cui:12},\cite{Yi:12},\cite{deng2012scintillation}. A common thread of all the works cited below and the work that followed is the prominence of analytical methods as opposed to numerical methods like WOS in exploring phenomenology.  

While Martin and Flatt\`{e} did use WOS to examine the effect of random media on intensity scintillation it lacks a connection to modern atmospheric propagation work. It is also useful to understand the limitation of WOS and the effect of a change of power-law on those limits. In an earlier work \cite{Grulke} we made an early attempt at making these connections and understanding the relationship between the power-law exponent of a random media and scintillation as function of turbulence strength. Our conclusions, though, were that an explicit inner and outer scale were necessary. 

In this work, we describe the results of a comprehensive WOS campaign, started in \cite{bos2015simulation} for plane waves propagating in uniform non-Kolmogorov turbulence volumes. Results are presented here for intensity scintillation as a function both Rytov number and power-law exponent. Results are limited to a single propagation geometry featuring a finite inner and outer scale. Recent results have shown that failure to account for these features can result in inaccuracies when evaluating some quantities\cite{beck2022wave}\cite{beck2021saturation}. We find that peak Normalized Intensity Variance (NIV) increases with power-law. Similarly, the Rytov number where peak NIV occurs also increases. This confirms earlier results \cite{bos2015simulation} indicating a shift of the NIV curve up and to the right as power-law increases. Across the campaign WOS are run to their practical limit for this propagation geometry assuming an upper bound of 8192 x 8192 phase screen samples. Thus, this work also describes the upper bound of WOS as a function of non-Kolmogorov power-law using Martin and Flatt\'e’s sampling constraints. Interesting here is that smaller-power laws are limited to Rytov numbers of about 7 compared to 12 for Kolmogorov turbulence and larger power-laws exponents.

The remainder of this paper is outlined as follows, following this introduction we provide an overview of WOS model and the sampling constraints used to define our simulation campaign. In section \ref{sec:Methods} we describe the campaign in detail and describe our scoring quantities. Results are provided in Section  \ref{sec:Results} followed by conclusions and directions for future work in Section \ref{sec:Conclusions}. 
\section{BACKGROUND} 
\label{sec:back}
A goal of this work is to understand the practical limits of WOS via phase screens generated by filtering white Gaussian noise by the Power Spectral Density (PSD) spectrum of the turbulence fluctuations and using the split-step propagation method. Excellent descriptions of the split-step propagation technique can be found elsewhere\cite{voelz2011computational},\cite{schmidt2010numerical} and are not included here. 

In approaching this work, we use the sampling constraints outlined by Martin and Flatt\'e\cite{martin1988intensity} for plane waves to ensure that the constraints account a non-Kolmogorov power-law medium. This limit is based upon the spatial bandwidth of the power-spectrum as sampled by the phase screen. As pointed out indirectly by Martin and Flatt\'e and elsewhere the power-law affects the prominence of high versus low frequency fluctuations in the media. For this reason, power-law random media with smaller power law are likely to have more high frequency fluctuations and therefore require a higher sampling rate compared to the Kolmogorov default. 
 Before describing those constraints let us first define the three-dimensional PSD of index of refraction fluctuations for a generalized turbulence volume described by a power-law with an inner and outer scale.
 \begin{equation}
\Phi_{n}(\kappa, \alpha, z) = A(\alpha)  \beta(z) \exp( -\kappa^2 / \kappa_{m}^2) (\kappa^2 + \kappa_{0}^2)^{(-\alpha/2)}
\label{eq:nokSpectrum}
 \end{equation}
 In Eq.\ref{eq:nokSpectrum} $A(\alpha )=(1/4{{\pi }^{2}})\cos \left( \frac{\pi \alpha }{2} \right)\Gamma [\alpha -1]$. In this description, the power-law exponent, $\alpha$, is restricted to values $3 < \alpha < 4$. The term $\beta(z)$ is a stand-in for the index of refraction structure constant, $C_{n}^2$ and has units of $m^{3-\alpha}$. The term $\kappa_m = c_1(\alpha)/l_0$, represents the inner scale of turbulence, $l_0$, and is defined as
${{c}_{1}}(\alpha )=2{{\left( \frac{8}{\alpha -2}\Gamma \left[ \frac{2}{\alpha -2} \right] \right)}^{\frac{\alpha -2}{2}}}$
In Kolmogorov turbulence, $\alpha = 11/3$, $A(\alpha) = 0.033$, $c_1(\alpha) = 6.88$. The term $\kappa_0$ sets the outer scale of turbulence, $L_0$, and is either set directly as $\kappa_0 = 1/L_0$ or as $\kappa = 2\pi/L_0$ and Eq.\ref{eq:nokSpectrum} reduces to the modified von Karman spectral model. 

It is common for WOS to use the Fried parameter, $r_0$, to set turbulence strength for phase screens. In a previous work\cite{bos2016simulation}, one of us compared two methods of setting the phase screen turbulence strength for studies of non-Kolmogorov turbulence. One method used the non-Kolmogorov equivalent Fried parameter so that each screen has the same effective spatial coherence properties. The other technique relied on first normalizing the spectral energy and then scaling the energy in the screen by the equivalent fluctuation energy that would be found in a Kolmogorov turbulence volume. A finding of this work \cite{bos2016simulation} was that the latter method provides a better match to ERT when evaluating scintillation index in terms of the plane wave Rytov number ${{\sigma }_{R}}={{(1.23C_{n}^{2}{{k}^{7/6}}{{L}^{11/6}})}^{1/2}}$ where $k$ is the optical wavenumber and $L$ is the propagation distance.

In this work, phase screens are generated using the technique described originally in \cite{bos2015anisotropic} and extended to WavePy\cite{beck2016wavepy}. The current version of WavePy uses the sub-harmonic method described by Johansson and Gavel\cite{johansson1994simulation}. In a previous work\cite{bos2015anisotropic} one of us noted a practical limit of around $\alpha = 3.8$  for accurately modeling non-Kolmogorov phase screens with an infinite outer scale. Anecdotally, we find that this new method improves phase screen accuracy in terms of a structure function match and allows us to explore out to $\alpha = 3.9$. Though, as we will discuss later as $\alpha \to 4$ the turbulent disturbance becomes equivalent to a pure tilt \cite{fried1965statistics}. It follows then that improved subharmonic modeling would allow for larger power-laws to be modeled more accurately. 

In determining sampling rates for our WOS campaign we use the spatial bandwidth definition described by Martin and Flatt\'e  to determine the required spatial bandwidth, ${R}_{\chi }$ , in each phase screen as 
\begin{equation}
{{R}_{\chi }}={{2}^{\left( 1-\frac{4}{\alpha }+\frac{1}{2+\alpha } \right)}}{{5}^{\left( \frac{1}{2+\alpha } \right)}}{{\sigma }^{2/\alpha }}{{\left( \frac{\left( 2+\alpha  \right)\sec \left[ \frac{\pi \alpha }{4} \right]}{\alpha \Gamma \left[ 1+\frac{\alpha }{2} \right]} \right)}^{2/\alpha }} %
\label{eq:spatialBW}
\end{equation}
here 
\begin{equation}
{{\sigma }_{R}}^{2}=C(\alpha )\beta {{k}^{1-\alpha /2}}{{L}^{2+\alpha /2}}    
\label{eq:sigmaNOK}
\end{equation}
and 
\begin{equation}
C(\alpha )=-\frac{2\Gamma \left[ -\frac{\alpha }{2} \right]\Gamma \left[ \alpha +3 \right]}{\left( \alpha +2 \right)\left( \alpha /2+1 \right)}\cos \left( \frac{\pi \alpha }{2} \right)\sin \left( \frac{\pi \alpha }{2} \right)
\end{equation}
In Eq.\ref{eq:spatialBW} power spectrum index has bounds $3<\alpha <4$. Note that we have used the more common 2D isotropic power spectrum representation as opposed to the 1D representation used by Martin and Flatt\'e where $1<\alpha <2$ and have adjusted the expressions presented here accordingly. From Eq.\ref{eq:spatialBW} we can find the required spatial sampling as $\Delta x=1/2{{R}_{\chi }}$. For the single simulation scenario considered in this work we fix the side length, $D = 1$ m,  so that the required number of samples for each phase screen is $N=1/\Delta x=2{{R}_{\chi }}$.

Another requirement described by Martin and Flatt\'e \cite{martin1988intensity} and others\cite{schmidt2010numerical} limits the log-amplitude variance to not exceed 0.1 or 10\% between propagation steps. This is effectively a forward-scattering requirement or the Rytov approximation and can be set in terms of $\alpha$  as 
\begin{equation}
\Delta L=\sqrt[1+\alpha /2]{C(\alpha )\beta {{k}^{2-\alpha /2}}{{\sigma }_{R}}^{2}}
\label{eq:nScreen}
\end{equation}
With the consequence that the minimum number of screens for a fixed propagation distance, $L$, is 
${{n}_{scr}}= \textnormal{ceil}\left( \frac{L}{\Delta L} \right)+1$
Having laid out these requirements we can now define a set of propagation parameters for our fixed scenario laid out in the next section.
\section{Methods}
\label{sec:Methods}
The main result of this work is an extensive WOS campaign that evaluates the scintillation index or NIV as a function of the plane wave Rytov number in non-Kolmogorov turbulence. Here we intend to hold the path length fixed and vary only the turbulence strength of the volume in terms of the plane-wave Rytov number, $\sigma_R$, with the goal of performing simulations out to the limit allowed by the sampling constraints described in Section \ref{sec:back}. For the purposes of this work the maximum number of samples per screen will be $2^{13} = 8192$ square. While $2^{14} = 16384$ is feasible, for the scenarios here that sampling is required only for conditions well-within the saturation region and where wall-clock executions times are considerable. 

The WOS campaign was run in two batches, the first was completed out to the maximum Rytov number allowed for one of three power laws, $\alpha$. Power laws values of $3.1$ and $3.9$ cover the lower and upper limits of practical simulation, while the Kolmogorov power value of $11/3$ $(3.66)$ was included as a baseline. A second, more limited, campaign was conducted for twenty-five values of $\alpha$ between $3$ and $4$. Values started at $\alpha = 3.02$ and increased increments of $0.02$ up to a value of $\alpha = 3.2$. From there $\alpha$ was incremented in steps of $0.1$ up to $3.8$ over the regime including the Kolmogorov value ($\alpha = 3.66$) included as a special case. From $\alpha = 3.8$ to $3.98$ steps were reduced again to $0.02$. The objective here being to identify values of power-law exponent where the simulation is no longer valid, or fidelity is reduced. 

As mentioned in the previous section we fixed the side length $D =1$ m and varied the sampling rate $N$ as required. For the other WOS parameters, the wavelength is set to $\lambda = 1$ $\mu$m, and the path length to $L = 5$ km. For Kolmogorov turbulence these parameters allow for a full exploration of the various propagation regimes described by theory without resorting to values of $C_{n}^2$ unlikely to be measured in the field; in the range of $10^{-17}$ to $10^{-12}$ $m^{-2/3}$. 

Using the Eqns.\ref{eq:sigmaNOK} and \ref{eq:nScreen} in Section \ref{sec:back} we are able to evaluate the maximum Rytov number that can be evaluated for a given sampling rate, $N$.  Simultaneously, for a given value of $\alpha$, we can find the minimum sampling rate needed to evaluate a specific Rytov number to ensure accurate results. As outlined in \cite{Grulke} it is possible to model atmospheres with power-laws very near to the lower-bound of $\alpha=3$ via WOS. However, as the value of $\alpha$ approaches $3$ the number of samples required to simulate even weak turbulence is very large. This both unnecessarily limits the range of volume turbulence strengths we can explore and increases wall-clock simulation time. Using a value of $\alpha = 3.1$ as our lower-limit case provides some relief in these regards while allowing a detailed examination of the behavior at smaller power law exponents. 

In our previous, related works \cite{beck2022wave},\cite{beck2021saturation}, we found that it is likely that WOS simulations without finite inner, $l_0$, and outer, $L_0$, scales are likely not valid. For this work, the outer scale size was set to $L_0=1$ m and the inner scale to $l_0 = 0.005$m. The former matches the fixed screen size, $D$, while the latter is slightly larger than the largest spatial sampling rate of $D/N$ for $N=256$ of $\Delta x = 0.0039$ m.

\subsection{Limits on WOS in non-Kolmogorov turbulence}
\begin{figure}
\begin{center}
\begin{tabular}{cc}
\includegraphics[width = 0.4\columnwidth]{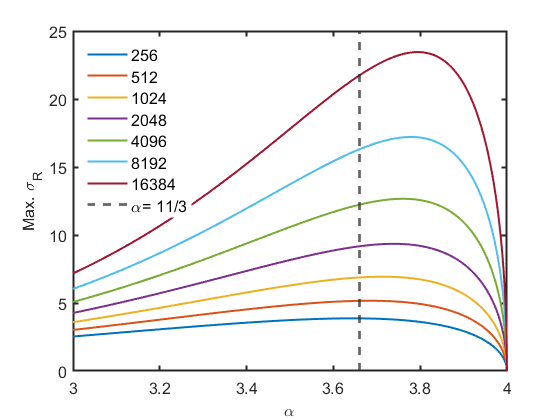}& \includegraphics[width = 0.4\columnwidth]{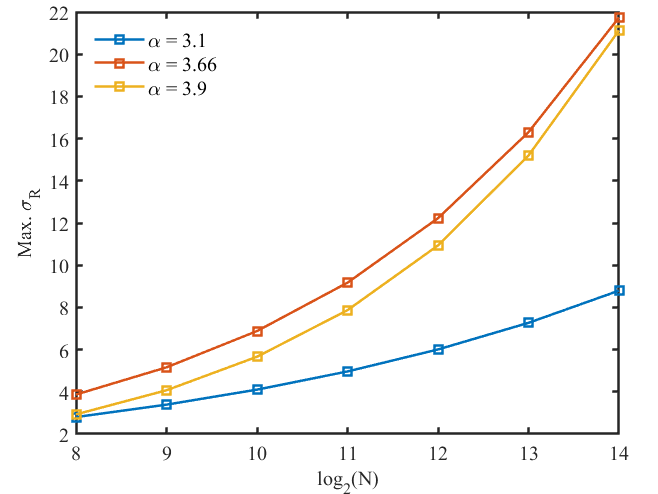}\\
(a) & (b)
\end{tabular}
\end{center}
\caption[Limits on WOS in NOK turbulence]{\label{fig:Limits} Limits on the maximum plane wave Rytov number, $\sigma_R$ that can be simulated for a specific power-law exponent, $\alpha$ (a) or screen size, N (b). The three values of $\alpha$ indicated are those chosen for detailed evaluation.}
\end{figure}
In Fig.\ref{fig:Limits} we evaluate Eq.\ref{eq:sigmaNOK} and \ref{eq:nScreen} in order to visualize the limits on WOS in non-Kolmogorov turbulence as described in this paper. In subfigure (a) the maximum Rytov number allowable for each screen size of $N=2^n$ where $n = 1,2,..14$ is presented. Of note here is that the maximum Rytov number that can be simulated with a screen size of $N=2^{14}$ is close to $\sigma_R = 24$. Also, that for all values of sampling rate the power-law allowing the highest turbulence strength is close to $\alpha=3.8$ it is also at this point where additional sampling most increases the range of Rytov numbers covered. While as $\alpha \to 3$ the same benefit is not conveyed, and the maximum turbulence strength is $\sigma_R = 6$ for $N=2^{13}$ and $\sigma_R = 7$ for $N=2^{14}$. Though not captured in our campaign we also note that as $\alpha \to 4$ the max Rytov number drops to zero. Again, this is because media with a power law of exactly $4$ are pure tilts and cannot be described via a structure function. This also explain the difficulty observed in \cite{bos2012technique} of generating accurate phase screen statistics at larger values of $\alpha$ and further underscores the practical limit of $\alpha = 3.9$ for WOS modeling. 
\subsection{Simulation Banding}
\begin{table}
\begin{center}
\begin{tabular}{|c|c|c|c|c|c|c|}
\hline 
N / $\alpha$ & 256 & 512 & 1024 & 2048 & 4096 & 8192 \\ 
\hline 
3.9 & 0-1.6 (10) &1.8-2.8 (15) & 2.9-4 (20) & 5 (20) & 6,7 (25) & 8,9 (35), 10-12 (60) \\ 
\hline 
3.66 & 0-1.8 (10) & 2 - 3 (15), 3-4 (25) & 4,5 (25) & 6 (30) & 7-9 (40) & 10-12 (60) \\ 
\hline 
3.1 & 0-2.6 (15) & 2.6 - 3 (20) & 3.2-3.8 (25) & 4,5 (35) & 6 (60) & 7 (60) \\ 
\hline 
\end{tabular} 
\caption{\label{tab:banding} Range of turbulence strengths $\sigma_R$ for each set of screen sizes, $N$, and ($n_{scr}$) per propagation. For each $\sigma_R$, 100 turbulence volumes were modeled for most cases. In the focusing regions 200 runs were performed in some instances. }
\end{center}
\end{table}
The figure on the right in Fig. \ref{fig:Limits} shows maximum $\sigma_R$ for the three values of $\alpha$ considered in the first batch of simulations as a function of sampling rate. In Table \ref{tab:banding} we lay out the banding for the first batch of our WOS campaign. For each value of $\alpha$ and $N$ the range of Rytov numbers is indicated. The number of WOS steps is indicated in parentheses. In the region approximately bound by $1.2 < \sigma_R < 2.$ for $\alpha = 3.66, 3.9$ a total of $200$ Monte-Carlo runs were executed for each value of $\sigma_R$ specified. Runs in the weak turbulence regime and saturation region required only $100$ runs to acheive statistical convergence.
\subsection{Calculation of NIV}
As defined by Andrews and Phillips\cite{andrews2005laser} the NIV or scintillation index, $\sigma_{I}^2$, is defined as the normalized variation in intensity such that
\begin{equation}
\sigma_{I}^2 = \frac{<I^2>}{<I>^2}-1
\label{eq:scintIdx}
\end{equation}
where the angle brackets indicate an ensemble average. In our simulations here, we wish to evaluate the scintillation index in the receiver plane for a plane wave source. The outcome of each simulation Monte Carlo trial is an intensity distribution in the receiver plane. The question, then, is how and where to evaluate scintillation in each case. 

For a plane wave, defined as uniform in amplitude and infinite in extent, propagating in a vacuum the expected intensity profile in any receiver plane is similarly uniform in amplitude, and therefore, intensity. Consequently, we can evaluate the scintillation index at any point in the receiver plane over the ensemble of turbulence volumes or runs. Doing so increases the amount of averaging allowing evaluation of the fluctuation statistics with fewer Monte Carlo trials. Practically, though, we must account for the fact that our propagator \cite{beck2016wavepy} incorporates a super-Gaussian absorbing boundary to prevent wrap-around artifacts. For this reason, and to avoid the influence of wraparound or boundary effects we evaluate scintillation index for each point inside a circle with a radius of $D/4 = 0.25$m from on axis center of the receive plane.
\section{Results}
\label{sec:Results}
\begin{figure}
\begin{center}
\begin{tabular}{c}
\includegraphics[]{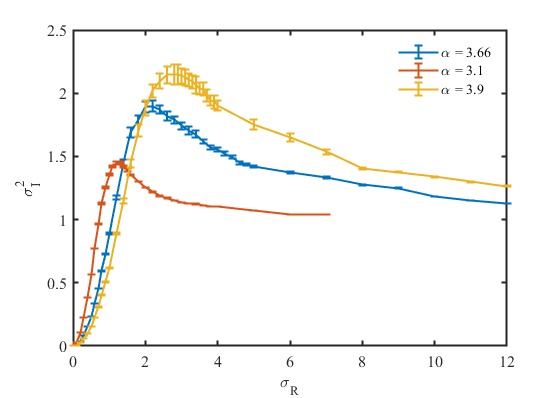}
\end{tabular}
\end{center}
\caption[Plane Wave Results Full Range or turbulence strength]{\label{fig:PWResults} Plane wave simulation of scintillation index, $\sigma_{I}^2$, as a function of Rytov number, $\sigma_R$, for $\alpha= 3.1, 3.66, 3.9$. For each power-law simulations were conducted out to the maximum Rytov number specified for a sampling rate of $N$=8192. In this scenario the pathlength was $5000$ m and the wavelength was $1$ $\mu$m. Error bars indicate the variance about the mean at each data point.For all values of $\alpha$ the outer-scale size was $L_0 = 1$ m and inner-scale was $l_0 = 5$ mm.}
\end{figure}

Fig.\ref{fig:PWResults} provides the results of the first of our two WOS campaigns. For each specified value of $\alpha$ scintillation as measured in our WOS is plotted out to the maximum Rytov number allowed according to the Eq.\ref{eq:spatialBW} for a screen with $N=8192$ samples. As described in Table \ref{tab:banding}, $\alpha = 3.1$ is limited to a maximum of $7$ while $\alpha = 3.66, 3.9$ go to $\sigma_R = 12$. The results here are similar to those presented elsewhere and previously \cite{bos2015simulation},\cite{Grulke},\cite{toselli2007scintillation} but in this instance include an explicit inner and outer scale in the turbulence spectrum sized based on the simulation geometry.

Examining Fig. \ref{fig:PWResults} we see the that the curves shift up and to the right relative to power-law. Earlier onset of saturation for the smaller power law case is also observed though without the sharp peak in the transition in the focusing region observed in \cite{bos2015simulation}. We attribute this difference to the explicit, finite, inner-scale in the turbulence spectrum in Eq.\ref{eq:nokSpectrum}. This spectral feature filters, or limits, the degree of small-scale fluctuations that drive scintillation and loss of spatial coherence resulting in saturation. Note that even if an inner scale is not an explicit feature of the turbulence power spectrum one is included, implicitly, by the simulation sampling rate. In WOS campaigns, like ours, where the sampling rate increases with turbulence strength, the inner scale would decrease with each change in sampling rate. Analytic theory indicates \cite{andrews2005laser}that this change should hasten onset of saturation and reduce peak scintillation. Here, our chosen value of $l_0 = 5$ mm is larger than the largest sampling rate, $\Delta x$, used in the WOS to avoid this complication.

In some previous works\cite{bos2015simulation},\cite{toselli2007scintillation} it was noted that scintillation may increase without bound as $\alpha \to 4$, and also that this behavior is missing in Fig.\ref{fig:PWResults}. Here, also, the inclusion of an explicit finite outer scale likely plays a role by limiting the relative energy in large-scale fluctuations\cite{Yi:12}. However, the observation holds that, all things being equal, peak scintillation increases with power-law and occurs at higher equivalent turbulence strength. So that, relative to Kolmogorov turbulence, an increase power law increases peak scintillation and results in relatively higher scintillation in the saturation region. Conversely, if the power-law of the medium is decreased peak focusing scintillation and scintillation in the saturation region are smaller.

On the other hand, for plane waves propagating in weak-to-moderate strength turbulence volumes ($\sigma_R < 1$) the situation is reversed. In this regime the relationship between turbulence strength and intensity scintillation is driven by the power-law of the medium. Consequently, we observe that mediums with smaller power laws experience higher fluctuations in intensity. Likewise, mediums with smaller power-laws experience less focusing and enter saturation sooner as a function of turbulence strength. 
\begin{figure}
\begin{center}
\begin{tabular}{cc}
\includegraphics[width = 0.4\columnwidth]{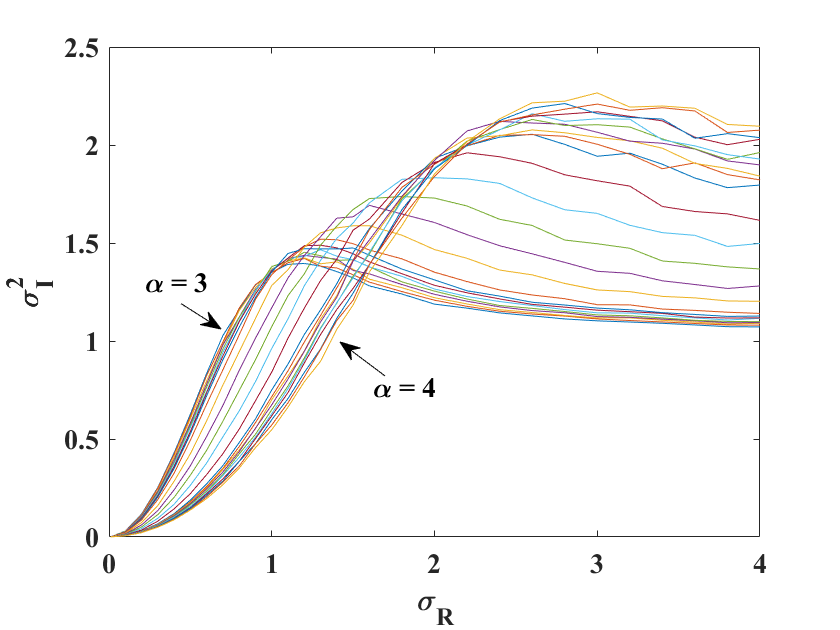}& \includegraphics[width = 0.4\columnwidth]{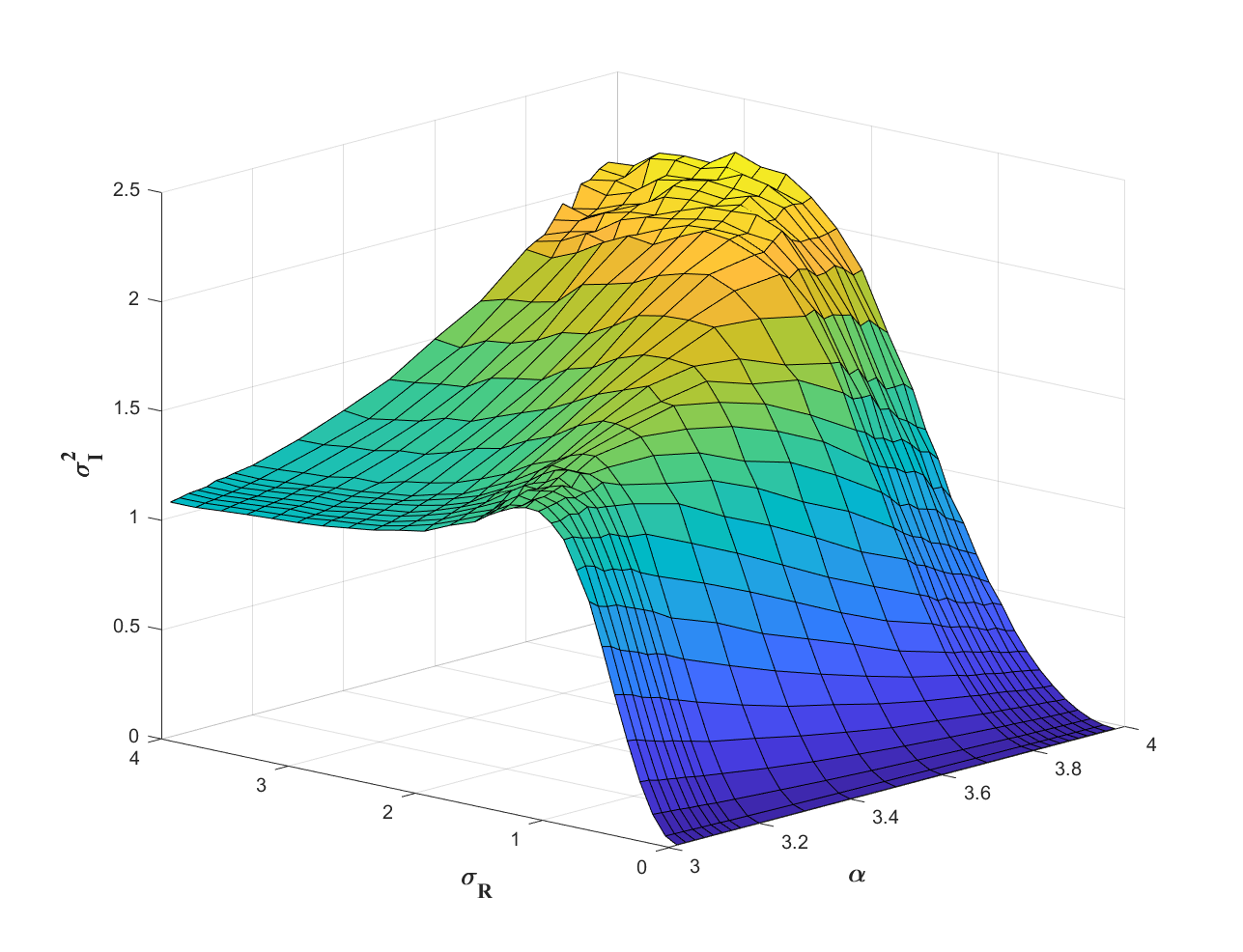}\\
(a) & (b)
\end{tabular}
\end{center}
\caption[Plane Wave Results full range of alpha]{\label{fig:PWAlpha} (a) result of WOS for $25$ values of $\alpha$ in the range $3 < \alpha$ <4 between $0 < \sigma_R < 4$. (b) surface plot of (a).}
\end{figure}

The results of the second batch of our WOS campaign are presented in Fig.\ref{fig:PWAlpha} and confirm these results further. All parameters are the same those in Fig.\ref{fig:PWResults} but the value of $\sigma_R$ is limited to a maximum of $4$ while $\alpha$ was varied as described in Section \ref{sec:back}. Subfigure (a) can be compared directly to Fig.\ref{fig:PWResults} and confirms that the behavior observed there is approximately continuous and increases monotonically as a function of $\alpha$.
\begin{figure}
\begin{center}
\begin{tabular}{cc}
\includegraphics[width = 0.4\columnwidth]{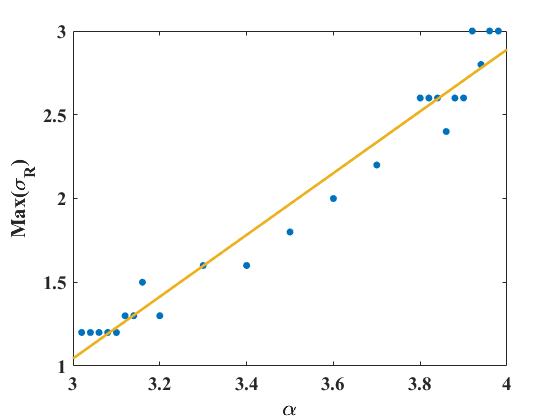}& \includegraphics[width = 0.4\columnwidth]{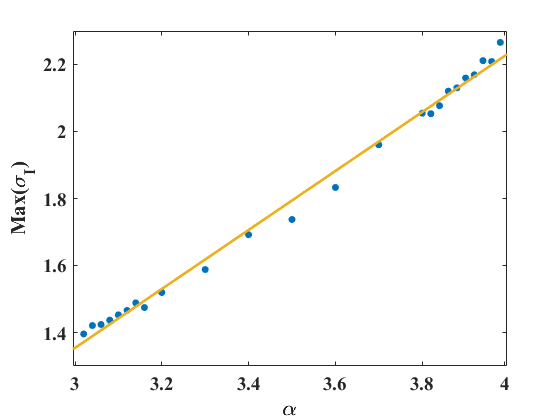}\\
(a) & (b)
\end{tabular}
\end{center}
\caption[Scintillation as a function of $\alpha$]{\label{fig:trends}. Value of $\sigma_R$ where maximum scintillation occurs (a) and peak value of $\sigma_I^2$ (b) as a function of power-law index $\alpha$}
\end{figure}

In Fig.\ref{fig:trends} the peak value of scintillation index (right) and the plane wave Rytov number where the peak scintillations were observed (left) are plotted as power-law, $\alpha$, is varied. In both figures, the least squares linear fit to the data is also plotted. The relationship between $\alpha$ and maximum scintillation is observed to have a slope of $0.9$ while the slope of the Rytov number where peak scintillation occurs and $\alpha$ is $1.8$. In this latter case, values necessarily are restricted to the specific values of Rytov number simulated during the campaign resulting in binning or quantization errors. In both cases, as we noted earlier, WOS accuracy is not certain as we exceed the bounds of $3.1 < \alpha < 3.9$ also contributing some uncertainty to these results. Regardless, the evidence of a positive linear relationship between power-law and both peak intensity scintillation and the volume turbulence strength at which that peak scintillation occurs is clear. For this propagation scenario it is also clear that the turbulence value of peak scintillation moves to the right at about double the rate of the peak scintillation increases. 

To our knowledge, this linear relationship between power-law of the medium and peak scintillation in terms of Rytov number has not been reported elsewhere. Toselli\cite{toselli2007scintillation} used ERT to generate analytical models of plane wave scintillation as a function of power-law and Rytov number. However, this model does not include finite inner and outer scales. Though, that work is consistent with these findings and our previous work \cite{bos2015simulation}\cite{bos2016simulation}\cite{Grulke}. In subsequent works by Toselli \cite{toselli2009free} and others \cite{deng2012scintillation},\cite{Yi:12}, \cite{Cang:11}, \cite{Cui:12},\cite{YI2013199} on non-Kolmogorov turbulence findings are often reported for a fixed value of $\beta$ in the weak to moderate regime while varying propagation distance instead of normalized volume turbulence strength as is done here. As pointed out by Charnotskii \cite{charnotskii2011twelve} because the units of $\beta$, or $\tilde{C}_{n}^2$ in some other works, vary with $\alpha$ it is not possible to use a common length scale for comparing results in any meaningful way. As far as we can ascertain, all previous works generally report a relationship similar to \cite{toselli2009free} where peak scintillation occurs at a $\alpha < 11/3$ in the region of $3.2 < \alpha < 3.3$ and is lower on either side of the peak going to zero as $\alpha \to 3$ and to a small value as $\alpha \to 4$.
\begin{figure}
\begin{center}
\includegraphics[]{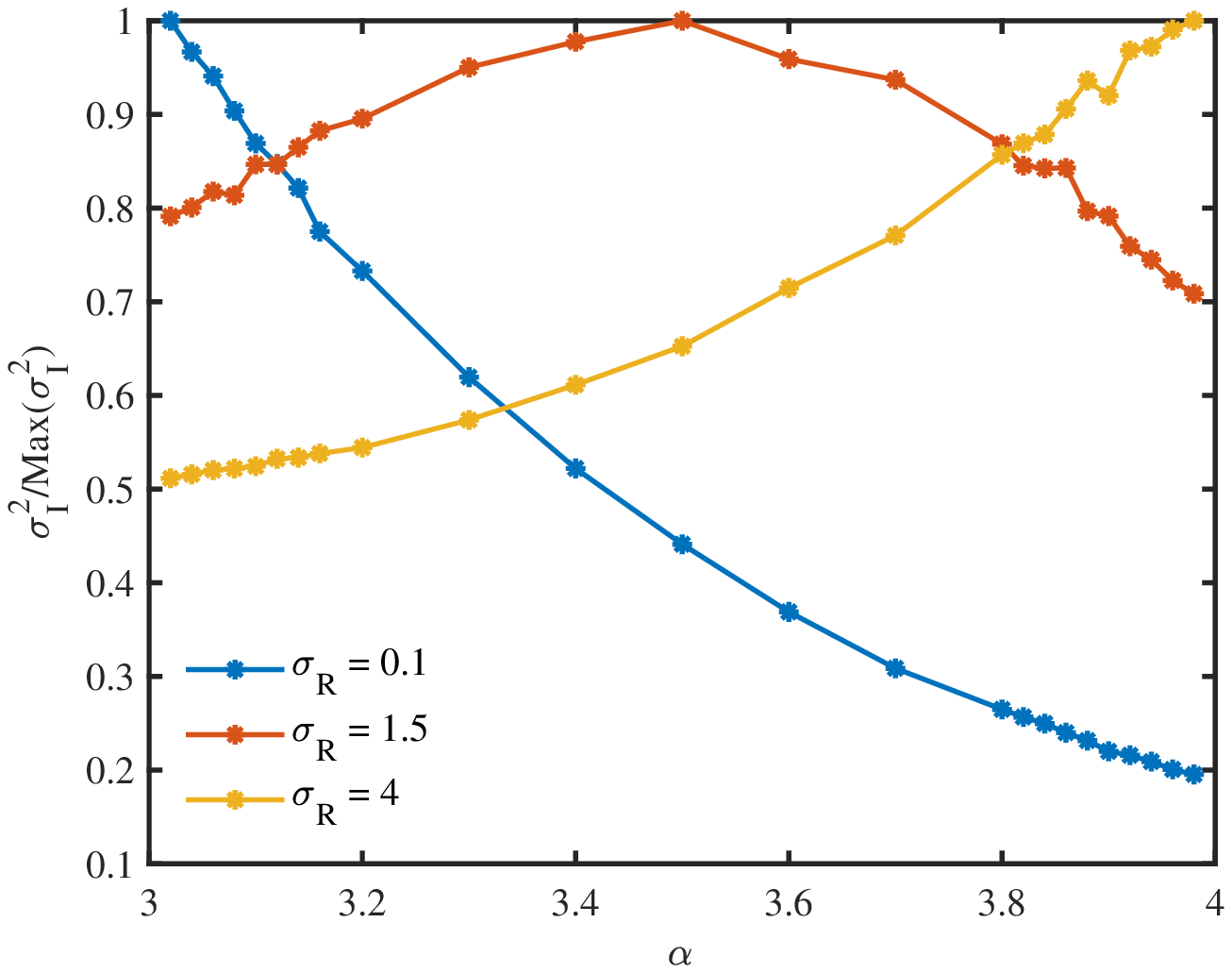}
\end{center}
\caption[Scintillation for fixed $\sigma_R$]{\label{fig:fixed}. Scintillation index for three fixed values of integrated turbulence strength in terms of $\sigma_R$ as a function of $\alpha$. Each trace is normalized to the maximum value of scintillation index over all values of $\alpha$}
\end{figure}

To place our work in this context, in Fig.\ref{fig:fixed}, we have plotted the normalized scintillation index, $\sigma_{I}^2/Max(\sigma_{I,\alpha})^2$, as a function of power-law index, $\alpha$. As pointed out earlier, we see that in the weak turbulence regime, for a fixed volume turbulence strength, scintillation decreases with power-law. In contrast, scintillation increases if the volume is characterized as strong or deep ($\sigma_R >> 1$). However, in this figure when $\sigma_R = 1.5$ scintillation peaks near the center of the range at $\alpha  = 3.5$ and decreases if the power-law is increased or decreased. 

Thus, relative to those other works the relationship between $\alpha$ and $\sigma_{I}^2$ varies with integrated turbulence strength in the volume. Also, because of the ambiguity introduced via a fixed $C_{n}^2$ this behavior may change depending on the power-law so that any of the behaviors observed in Fig. \ref{fig:fixed} may apply. For example, if $\lambda = 1.55$ $\mu m$, $L = 1000$ m, and $\beta = 7 \times 10^{-14}$ m$^{3-\alpha}$, as in \cite{toselli2009free}, the Rytov number varies from $3.8$ to $1.18$ to $0.73$ for $\alpha = 3.1, 3.66,$ and $3.9$. Consider then that for this one example all three of the behaviors in Fig. \ref{fig:fixed} may be observed. This observation further emphasizes that some measure of equivalent volume turbulence strength must be when trying to understand the impact of turbulence power-law exponent on beam propagation and imaging. 
\section{Conclusions and Future Work}
\label{sec:Conclusions}
In this work we explored the limits of WOS plane-waves propagating in non-Kolmogorov turbulence and with power-law exponents in the range $3 < \alpha < 4$. At the upper bound the medium becomes a pure tilt and therefore WOS are limited by the size of the screen or the fidelity of the low-spatial frequency compensation mechanism; sub-harmonics for example. Conversely, as the power-law approaches the lower-bound the medium become spatially uncorrelated, and the spatial bandwidth required to properly simulate the medium becomes very large. Accordingly, the number of samples required by the simulation is also large. Both WOS requirements, can be ameliorated by the inclusion of a finite inner and outer scale. Indeed, as we have shown elsewhere\cite{beck2021saturation} if these values are not defined explicitly in the power-spectrum model they are implicit in the simulation model. Finally, if we take the practical limit of WOS to limited to $N = 16384$ samples, the maximum plane wave Rytov number that can be simulated in any power-law medium is $\sigma = 24$ when $\alpha = 3.8$. 

The results of our analysis on the limits of WOS were used to inform two simulation campaigns exploring the interplay between Rytov number, $\alpha$, and scintillation of intensity or NIV. The first campaign aimed to evaluate scintillation as a function of Rytov number for values of power law $\alpha = 3.1, 3.66, 3.9$. For each value of $\alpha$ simulations were undertaken out the maximum $\sigma_R$ allowed for a screen size of $N = 8192$. Consistent with other results the maximum scintillation observed and the Rytov number where the peak occurs increase with power-law. At small power-law exponents scintillation quickly saturates as the volume turbulence strength increases. In this work we did not observe a strong focusing peak seen in our previous works as $\alpha \to 3$. We attribute this to the use of a finite inner scale larger than the sampling rate. Similarly, in previous works scintillation appeared to increase without bound at power-laws near $\alpha = 4$. However, if we include an explicit outer scale on the order of the screen size peak scintillation increases but eventually rolls off into saturation. These findings are consistent with theory that attributes small-scale fluctuations to scintillation strength and large-scale fluctuations to the peak of intensity fluctuations in the focusing region. 

Our second simulation campaign aimed to empirically evaluate the effect of medium power-law index on scintillation. We found here that, in contrast to some previous works, intensity scintillation increases monotonically with power law. Likewise, the $\sigma_R$ where peak scintillation occurs moves to the right at twice the rate peak scintillation increases. We assert that the differences between these results and previous works is due to the proper scaling of turbulence strength with power-law used here. The aforementioned inner and outer-scale may also be contributing factors. 

Over the course of this work, the contradictory nature of our findings and previous works led us to revisit our results to ensure our WOS parameters were correct. As a result of these explorations\cite{Grulke}, we were led to conclude that, at least for the simulation scenario explored here, these parameters are a mostly a second order effect. That is to say, the overall trends remain as long even if the sampling rate or number of screens changes. This finding has been recently confirmed by Wijerathna \cite{wijerathna2021numerical}. In one of his last works Flatte\cite{flatte2000irradiance} indicated that WOS are an exact solution to the stochastic Hemholtz equation over all bounds when properly configured. Much has been made of the qualification but based on this work and others we feel WOS will usually provide accurate results even if they are not precise \cite{Grulke}. 

There is more work to be done in this area. For example, it may be worthwhile to further explore Fresnel zone effects by varying wavelength and path length. This work looked only at plane-wave propagation. Therefore, it may be interesting to see if further accommodations\cite{schmidt2010numerical} are needed to account for beam expansion and beam-wander for as diverging sources. This work also explicitly does not include a Hill ``bump'' or other features in the dissipation range. This exclusion is purposeful as it is not at all clear the nature of this feature when the medium is non-Kolmogorov. Finally, a straight-forward extension would be to explore the effect of larger values of $l_0$ and smaller values of $L_0$ for different power-laws as $\alpha$ is varied. 

\acknowledgments
This material is based upon work supported by the Air Force Office of Scientific Research under award number FA9550-17-1-0201. Any opinions, finding, and conclusions or recommendations expressed in this material are those of the author and do not necessarily reflect the views of the United States Air Force. This work was began when the main author was a NRC Postdoctoral associate with the United States Air Force Research Laboratory under the supervision of Dr. Venkata (Rao) Gudimetla and he is credited for have first suggested the WOS campaigns described herein. 


\bibliography{article}   
\bibliographystyle{spiejour}   


\vspace{2ex}\noindent\textbf{Jeremy P. Bos} is an Associate Professor of Electrical and Computer Engineering at Michigan Technological University. Before joining Michigan Tech, Bos worked as a NRC Postdoctoral Fellow under the Research Associateship Program with the Air Force Research Lab in Kihei, Hawai’i. He received his PhD and BS from Michigan Tech in 2012 and 2000 respectively and his MS from Villanova University in 2003. Before returning to academia in 2009 Bos had 10 years of experience in the automotive and defense industry where he was licensed as a professional engineer. He is a senior member of SPIE and IEEE, and an author on over 100 scholarly contributions. His research interests are in the areas of imaging and light propagation in random media, signal processing, and sensor fusion. 

\vspace{1ex}
\noindent Biographies and photographs of the other authors are not available.
\listoffigures
\listoftables

\end{spacing}
\end{document}